\documentclass[doublecol,figures]{epl2} 

\usepackage{epsfig}
\usepackage{amsmath}
\usepackage{amssymb}
\usepackage{wrapfig}

\title{Stall force 
   of polymerizing microtubules and filament bundles}

\author{J. Krawczyk  \and J. Kierfeld }

\newcommand{\eb}{\varepsilon_l}
\newcommand{\Fs}{F_{\rm stall}}

\newcommand{\ko}{k_{\rm on}}
\newcommand{\kf}{k_{\rm off}}
\newcommand{\kso}{k^{\ast}_{\rm on}}
\newcommand{\ksf}{k^{\ast}_{\rm off}}
\newcommand{\vn}{{\vec n}}

\institute{                    
  Physics Department, TU Dortmund University, 44221 Dortmund, Germany, EU 
}
\pacs{87.16.Ka}{Filaments, microtubules, their networks, 
      and supramolecular assemblies}
\pacs{87.16.A-}{Theory, modeling, and simulations}
\pacs{87.15.rp}{Polymerization}

\abstract{
    We investigate stall force and  polymerization kinetics 
    of rigid protofilaments in a microtubule
    or interacting  filaments in  bundles   
    under an external load force in the framework of a 
   discrete growth model.
    We introduce the concecpt of polymerization cycles to describe 
    the stochastic growth kinetics, which allows us to  
    derive an exact expression for  the stall force.
    We find  that the stall force is 
    independent of  ensemble geometry and load distribution. 
    Furthermore, the stall force  is proportional to  the number 
    of filaments and  increases linearly with 
     the strength of  lateral filament interactions.  
    These results are corroborated by  simulations,
    which also show a  strong influence of ensemble geometry 
    on growth kinetics  below the stall force.
}

\begin{document}

\maketitle

\section{Introduction}

Polymerization of  cytoskeletal filaments
is essential for various cellular processes, such as
motility or the formation of cellular protrusions including
filopodia or  lamellipodia \cite{bray,mogilner_rev}.
Single polymerizing filaments can generate forces 
in the piconewton range, as has been demonstrated 
experimentally for microtubules (MTs) \cite{dogterom1997}. 
Such force generation mainly relies on the gain in chemical 
bonding energy upon monomer attachment \cite{Hill1987}. 
An opposing force slows down filament growth, 
which finally stops at the  stall force
representing the maximal polymerization force a filament 
can  generate. Therefore, the stall force 
is the essential quantity to characterize 
polymerization forces.

Cellular force generating structures such as filopodia 
are  made of polymerizing ensembles of interacting 
actin filaments \cite{faix06}. 
Particularly important are bundles of parallel filaments, which 
can  hold together by  crosslinking proteins or unspecific 
attractive interactions. 
Stall forces of  polymerizing actin bundles could be determined
experimentally only recently \cite{footer07}. 

MTs are tubular filaments, which also  consist
 of an ensemble of typically 13  interacting protofilaments (PFs). 
The force velocity-relation  of polymerizing 
MTs has been experimentally determined in refs.\ 
\cite{dogterom1997,Janson2004}, where  stall forces around
5 pN have been obtained.

An ensemble of many non-interacting  filaments or PFs 
is  believed to have higher stall forces than a single 
filament because of load sharing effects. 
First fits of the experimental data on MT growth
in ref.\  \cite{dogterom1997}
 were based on the assumption of load sharing  and application 
of ratchet models for a single
 rigid PF \cite{peskin1993a}.
An explicit  continuous model 
for $N$  rigid  PFs in a MT  under load
 resulted in  stall forces $\propto N^{1/2}$ \cite{ostermogilner1999}.
For an analogous  discrete growth model of 
 PFs in a MT
 it has been shown  that the 
stall force  of  $N$ PFs increases  $\propto N$ 
 compared to a single PF \cite{vanDoorn2000}, 
in agreement with equal load sharing. 
Variants of this model which allow a better fit of experimental 
data were discussed in  \cite{Kolomeisky2001}.

In addition to load sharing effects,  
crosslinking  or attractive lateral interactions within  
filament bundles or between PFs 
can allow the ensemble to generate 
even higher forces  by exploiting the additional interaction energy 
\cite{mahadevan00}. 
For flexible filaments, 
zipping  mechanisms for force generation can even 
rely exclusively on attractive interactions  \cite{kuehne09}.
These results 
suggest that the stall force of interacting filaments 
or PFs increases by the additional 
 interaction energy per length 
that a bundle gains upon assembly. 
Lateral interactions between PFs of a 
polymerizing MT have been considered in refs.\ 
\cite{Tanase2004, stukalin2004a, son2005}.
Also for MT growth,   approximative analytical results 
in refs.\ \cite{Tanase2004, stukalin2004a} suggest that the stall force 
of interacting PFs  increases by the additional 
 interaction energy per length 
that the MT gains upon assembly.

Apart from this progress, an  exact result  for the stall force 
could not be derived so far. 
Furthermore, 
the geometry of the  bundle or tube, i.e.\ the relative 
positioning of filaments or PFs in the 
ensemble, has an impact on 
the mechanics of monomer insertion under load  and on 
the lateral interactions, which are involved.

In this Letter, we investigate the combined effects  of 
 attractive filaments interactions and ensemble architecture 
on the  growth kinetics under a compressive force using 
a discrete growth model \cite{vanDoorn2000,Tanase2004,stukalin2004a}. 
Based on the concept of polymerization cycles we derive 
 an exact analytical result for the stall force.
This result shows that for the discrete model introduced in refs.\
 \cite{vanDoorn2000,Tanase2004,stukalin2004a} 
 the stall force is a universal quantity,
 which only depends 
on the polymerization energy gain 
and the interaction strength between filaments or PFs.
The stall force is independent of ensemble geometry and independent 
of the distribution of load force and interaction energy between 
attachment and detachment rates. 
The result also shows that the stall force increases $\propto N$, 
i.e.,  linearly in the number of filaments. 

Using stochastic simulations based on the Gillespie
algorithm 
we find that the growth kinetics below the stall force  depends 
sensitively on the strength of interactions between 
rigid filaments in a bundle or PFs in a MT
{\em and} 
 on the geometry of the bundle or tube. 
We find different shapes of the force-velocity 
relation as well as a complex non-monotonic dependence of 
the growth velocity on the relative filament positioning. 
This dependence is  very pronounced 
at low forces well below the stall force but vanishes upon 
approaching the stall force
resulting in 
a geometry independent  stall force.

 Our results are of particular interest with respect to the growth 
 kinetics of MTs, which usually contain 13 PFs.
 Our results imply that a two-start helical structure 
(with  a helical pitch of one tubulin dimer), 
which is often 
 assumed in modelling, has a distinct force-velocity relation 
but an identical stall force as the actual   
 three-start helix structure 
(with a helical pitch of three tubulin monomers) 
found 
 by electron microscopy \cite{nogales2001}.

\section{Model and simulation}

\begin{figure}
\begin{center}
 \epsfig{file=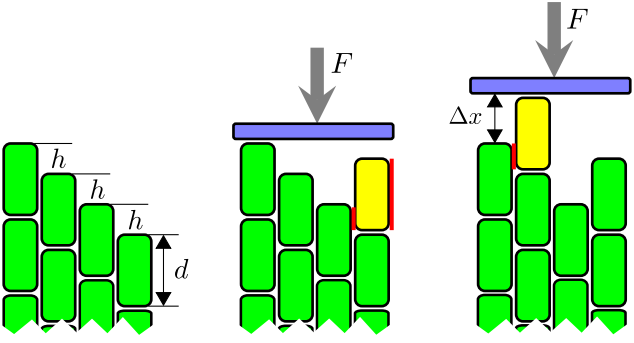,width=0.47\textwidth}
 \caption{ \label{pic_model}
    Schematic picture of the model of interacting PFs 
    polymerizing against the external load force $F$ for  $N=4$.
   Positions of
   monomers of size $d$  in neighboring PFs 
   are shifted by the offset $h$.  Insertion of 
   a monomer (yellow) might 
   change the position of the leading PF by  $\Delta x$ (right
  picture) and create
   additional contact length with neighboring PFs
    (red).
}
\end{center}
\end{figure}

We consider a filament consisting of $N$ rigid  PFs 
 in a tube-like arrangement such that 
each PF has two neighbors and periodic boundary 
conditions apply.
For actin bundles, this model neglects effects from thermal 
shape fluctuations \cite{KKL05} and the existence of defects within the 
bundle structure \cite{Gov2008,Yang2010, Grason2010}.
Each PF consists of monomers of size  $d$, see
fig.~\ref{pic_model}, 
The attachment and detachment rates for monomers 
are  $\ko$ and $\kf$, respectively, and related to the 
polymerization energy gain $E_p>0$ upon adding a monomer 
by $\ko/\kf = e^{E_p/k_BT}$ at temperature $T$. 
 For MTs, 
each monomer is a tubulin dimer, and 
we will neglect hydrolysis of GTP such 
that $\ko$ and $\kf$ are attachment and detachment rates for 
GTP-tubulin dimers.  
We also neglect catastrophes and consider MTs 
only in their growing phase. 
Effects of hydrolysis are shortly 
discussed in the end.

Each PF has  attractive lateral interactions with 
its two neighbors; the corresponding lateral association 
 energy (per length) is $\eb>0$. 
Thus, apart from the polymerization energy, an 
attaching  monomer gains   the lateral interaction energy 
$\eb\Delta \ell$, where  $\Delta \ell>0$ is  the additional 
contact length with
neighboring monomers, which is created upon inserting the monomer. 
Likewise, a detaching monomer looses a corresponding  interaction 
energy resulting in $\Delta \ell<0$. 
If the PF can equilibrate 
its remaining configurational degrees of freedom sufficiently 
fast during addition or removal of a monomer, 
thermodynamics requires that lateral 
interactions change on- and off-rates of that monomer  
such that 
$\kso/\ksf = (\ko/\kf) \exp(\eb|\Delta \ell|/k_BT)$.

The influence of an external load 
is described by a force $F$, which acts only 
on the leading  PF. We denote the position of the 
tip of the leading filament by $x$.  
Insertion of a monomer changes  the position of the tip 
 of the leading 
PF by $\Delta x>0$ in the on-process, 
which gives rise to an additional mechanical energy $F \Delta x$.
Likewise, removal of a monomer in the off-process gives rise to 
$\Delta x<0$.
If we assume again that the PF can equilibrate 
its remaining configurational degrees of freedom sufficiently 
fast during addition or removal of a monomer under force, 
thermodynamics requires  further  modification of 
 on- and off-rates such that \cite{peskin1993a,ostermogilner1999}
\begin{equation}
  \frac{\kso}{\ksf} = \frac{\ko}{\kf}
      e^{(\eb|\Delta \ell|-F|\Delta x|)/k_BT}.
\label{ratio}
\end{equation}

If on- and off-rates  $\kso$ and $\ksf$ are specified separately
the thermodynamic constraint (\ref{ratio}) allows to 
introduce  a load distribution factor 
$\theta_F$ and a lateral energy distribution factor $\theta_l$, 
\begin{eqnarray}
\kso &=&\ko      e^{ (-\theta_F F |\Delta x|+(1-\theta_l)\eb|\Delta \ell|)/k_BT}
\nonumber\\
\ksf &=& \kf     e^{ ((1-\theta_F) F |\Delta x| -\theta_l \eb|\Delta \ell|)/k_BT}.
\label{eq_onr1}
 \end{eqnarray}
In general,  load and energy distribution factors 
can depend on the specifics of insertion and removal and differ 
for each polymerization step.
A reasonable assumption is that
the external load only affects the on-rate 
(corresponding to $\theta_F=1$) 
and that the on-rate is diffusion-limited 
and not affected by lateral interactions
(corresponding to  $\theta_l=1$) but other 
choices are thermodynamically possible. 
In the absence of lateral interactions, this model (with $\theta_F=1$) 
was introduced by
van~Doorn~{\it  et~al.}~\cite{vanDoorn2000}.
Lateral interactions have been included  
in refs.\ \cite{Tanase2004, stukalin2004a, son2005}.

The geometry of the  bundle or tube has an impact on 
the mechanics of monomer insertion under load  and on 
the lateral interactions, which are involved. 
Therefore, we expect a strong influence of geometry 
on growth kinetics.
We control  the ensemble geometry by shifting the  relative position of 
neighboring PFs in the initial configuration by a distance 
$h$, which we call a {\it geometry parameter}, 
see fig.\ \ref{pic_model}. 
Because of the periodic boundary conditions,
 we have $N-1$ relative displacements of $h$ between neighboring 
PFs and one relative displacement of $(N-1)h$.
 For $h=0$ we have an aligned or  ``flat'' initial 
configuration with all PF tips at the same height. 
For symmetry reasons, 
geometry parameters $h$ and $-h$ are equivalent. The parameter 
$h$ can also be shifted by multiple monomer sizes $d$ 
corresponding to the insertion of additional monomers without 
changing the kinetics. Therefore, it is sufficient to consider 
 values $0<h<d/2$. 
In the following we will measure $h$ in units of $d$ and use
$\bar{h} \equiv h/d$.

Of special interest are MTs, which are usually 
built from $N=13$ PFs with tubulin dimers of size 
$d\simeq 8$ nm \cite{nogales2001}. 
Often it is assumed that MTs with $N=13$ PFs 
have an offset  $\bar{h} = 1/13$ \cite{vanDoorn2000,Kolomeisky2001},
which results in  a symmetric arrangement without a seam in the structure,
where the tube closes.
The actual structure exhibits 
a three-start helix 
with a helical pitch of three tubulin monomers (or $1.5d$) resulting in 
 $\bar{h} = 1.5/13$ \cite{nogales2001} and a seam.
Other numbers of PFs in MTs ranging 
from $N=8$ up to $N=19$ 
have been observed as well
\cite{chretien1991, nogales2001} in the form of 
two-start or four-start helices  corresponding to 
  $\bar{h} = 1/N$ (a helical pitch of $d$) or $\bar{h} = 2/N$ 
(a helical pitch of $2d$), respectively. 
In view of these different 
possible structures we want to study MT 
growth kinetics also 
as a function of the geometry parameter $h$ \cite{Tanase2004}.

In order to simulate the stochastic non-equilibrium 
growth dynamics of the model  we use the 
Gillespie algorithm \cite{gillespie1977}, which 
implements a continuous time Markov process with the 
rates introduced above. 
For the simulations we use parameters,
$\ko=200$ [1/min], $\kf=50$ [1/min], a  monomer size $d=8$ nm
 and a temperature $k_B T=4.1 {\rm pN}\,{\rm nm}$ 
corresponding to room temperature. 
We used a load distribution factor $\theta_F=1$ and performed 
simulations both for energy distribution factors $\theta_l=0$ 
(interaction energy affects on-rate)  and $\theta_l=1$ (on-rate
diffusion-limited). 
For each set of 
parameters, we average over $100$ runs.

\section{Polymerization cycles, growth velocity, and stall force}

The polymerization kinetics is completely 
determined by the configuration of the
$N$ PF {\em ends}: the absolute length 
of each PF does not enter the rates (\ref{eq_onr1})  
and we can neglect the possibility of vacancies or holes within a 
PF, which have not been observed experimentally.
Therefore, the    
 growth kinetics can be described by transitions between all 
possible states of the $N$ PF ends. 
In general, each such state of the filament end can be described by a 
set $\vn=(n_1,...,n_{N-1})$ of $N-1$ integer monomer 
number differences 
between neighboring PFs with 
$n_i$ as monomer number difference between PFs 
$i$ and $i+1$. 
Transitions between these states happen by monomer addition and 
removal with rates  
(\ref{eq_onr1}). Monomer addition (removal) 
at PF $i$ leads to changes 
$\Delta n_i=+1$ and $\Delta n_{i-1}=-1$ ($\Delta n_i=-1$ and $\Delta
n_{i-1}=+1$). The transition rates $k_{\vn_1,\vn_2}$ 
between two states $\vn_1$ and $\vn_2$ are only non-zero 
if both states are  related by  
addition or removal of a single monomer and the corresponding rates are 
determined by  (\ref{eq_onr1}), 
which depend on force and lateral interactions.

During polymer growth, layers of  monomers are added and eventually 
 layers with $N$ monomers  are  completed upon 
addition of a monomer. After addition of  $L$ complete  layers
the filament end attains the same configuration as initially. 
Therefore, 
each completion of $L$ layers closes a {\em polymerization cycle}
$C_L$ of transitions 
in the network of states $\vec{n}$.
A non-zero  average 
polymerization velocity  $v=\langle \dot x\rangle$ implies that 
layers are added with a non-zero rate and is equivalent to 
the existence of {\em stationary 
 cycle fluxes} in the network of states $\vec{n}$. 
We can calculate these  stationary cycle fluxes 
using general theorems derived for the kinetics
of chemical networks  \cite{hill1966,schnakenberg1976,hill1989}.

We denote the stationary 
 cycle flux for a polymerization cycle $C_L^+$ completing 
 $L$ layers by
$J(C_L^+)$. Likewise, the opposite cycle removing 
$L$ layers is called $C_L^-$ and the corresponding stationary 
flux is $J(C_L^-)$.
For an arbitrary  cycle 
$C_L^+=(\vn_1,\vn_2, ... , \vn_M,\vn_{M+1}\equiv \vn_1)$ 
of length $M$ and completing $L$ layers ($M\ge NL$), 
  the ratio of 
stationary cycle fluxes in forward and backward direction 
is given exactly by  the ratio of products of transition 
rates  
 along the edges of the cycles \cite{hill1966,hill1989}:
\begin{equation}
    \frac{J(C_L^+)}{J(C_L^-)}
    = \frac{ \prod_{i=1}^Mk_{\vn_{i},\vn_{i+1}}}{ \prod_{i=1}^Mk_{\vn_{i+1},\vn_{i}}}
\label{cyclefluxes}
\end{equation}

In addition, we can  establish 
 a  general relation between the average growth velocity 
$v$ and the stationary fluxes along cycles $C_L$: 
The total stationary net flux along the  fundamental set of all cycles 
$C_1$ for completions of {\em single} layers gives the
mean time to complete a single layer and, therefore, the mean growth 
velocity as  
\begin{equation}
    v = d \sum_{C_1}  
    (J(C_1^+)-J(C_1^-)) 
\label{v}
\end{equation}
Summation over 
single layer cycles $C_1$ is sufficient because they 
 form a fundamental set \cite{schnakenberg1976}, i.e., 
linear combinations allow to represent 
 cycles for an  arbitrary number of $L$ layers.

At the stall force, the growth velocity $v$ vanishes, i.e., 
$J(C_1^+)=J(C_1^-)$, for all polymerization 
cycles for single layers. Because such cycles form a fundamental 
set, it follows that all
stationary net polymerization cycle currents have to vanish, 
i.e., $J(C_L^+)=J(C_L^-)$ for {\em all} cycles 
$C_L$. This leads to the conclusion that the 
different filament end states are in detailed balance at 
the stall force, as has been conjectured in \cite{vanDoorn2000}. 
According to the relation (\ref{cyclefluxes}) 
we obtain the following  Wegscheider condition \cite{Wegscheider1901}
for any of the  polymerization cycles $C_L$ 
\begin{equation}
  \prod_{i=1}^M  \frac{k_{\vn_{i},\vn_{i+1}}}{k_{\vn_{i+1},\vn_{i}}} = 1.
\label{condstall}
\end{equation}
which will lead to an exact 
expression for the stall force.

We first consider the  cycle $C_L^+$.
Addition of exactly $L$ layers requires $M_+= (M+NL)/2$ attachment 
and $M_-=(M-NL)/2$ detachment transitions, where
$M=M_++M_-$  and $NL=M_+-M_-$. 
Regardless of load and energy distribution factors, the 
thermodynamic constraint~(\ref{ratio}) 
requires 
\begin{equation}
   \frac{k_{\vn_{i},\vn_{i+1}}}{k_{\vn_{i+1},\vn_{i}}}  
      = \frac{\ko}{\kf}
     e^{\pm(\eb|\Delta \ell_{i,i+1}|-F|\Delta x_{i,i+1}|)/k_BT}
\end{equation}
for each attachment (+) and detachment (-) transition in $C_L^+$.
Therefore 
\begin{equation}
  \frac{\prod_{i=1}^M k_{\vn_{i},\vn_{i+1}}}{\prod_{i=1}^M k_{\vn_{i+1},\vn_{i}}} = 
    \left(\frac{\ko}{\kf}    \right)^{M_+-M_-}
  e^{ (\eb|\Delta \ell_M|-F|\Delta x_M|)/k_BT   }
\end{equation}
where 
$\Delta \ell_M = \sum_{i=1}^{M_+} |\Delta \ell_{i,i+1}|-
    \sum_{i=1}^{M_-} |\Delta \ell_{i,i+1}|= LNd$
is  the total net  gain 
in lateral contact length
and 
$\Delta x_M = \sum_{i=1}^{M_+} |\Delta x_{i,i+1}|-
    \sum_{i=1}^{M_-} |\Delta x_{i,i+1}|= Ld$
is the total 
net advance of the leading PF.
As a result, we find for the  ratio of products of transition 
rates along the edges of the forward and backward cycles
the simple result
\begin{equation}
\frac{ \prod_{i=1}^Mk_{\vn_{i},\vn_{i+1}}}
    {\prod_{i=1}^Mk_{\vn_{i+1},\vn_{i}}} 
  = (\ko/\kf)^{LN} e^{(\eb LNd-F Ld)/k_BT}
\end{equation}
From the  condition (\ref{condstall}) 
we then   obtain  an {\em exact} expression for the stall force, 
\begin{equation}
 \Fs =N \left[ \eb + \frac{k_BT}{d} 
      \ln \left(\frac{\ko}{\kf}\right)\right].
 \label{Fs}
\end{equation}
Based on the assumption of detailed balance the same result has also 
been obtained in ref.\ \cite{Tanase2004}.
The stall force is a linear function of  the number $N$ of filaments in the 
ensemble and increases linearly with the lateral interaction.
 It is  independent  of all 
load or energy distribution factors of individual polymerization steps.
Moreover, 
the stall force  is {\em independent} of the ensemble architecture.
Our derivation shows that 
 geometry independence not only means that the result (\ref{Fs}) is   
independent of the parameter $h$ for an arrangement with 
constant offset between neighboring PFs
but that we obtain the same result (\ref{Fs})  for the 
stall force for completely arbitrary arrangements of PFs
relative to each other. Therefore, we also expect the 
stall force of a bundle of interacting actin filaments not to 
depend on the precise relative arrangement of actin filaments, 
which is hard to control in experiments.

In the framework of polymerization cycles 
the so-called one-layer approximation introduced in 
\cite{stukalin2004a} is equivalent to 
a ``one-cycle'' approximation, which 
restricts  the sum in (\ref{v}) to a contribution from a single 
cycle dominating the sum in the limit of large forces close 
to the stall force. Therefore, the exact expression (\ref{Fs}) for 
the stall force, at which all cycle currents become 
zero,  is also recovered  in the one-layer approximation in 
ref.\ \cite{stukalin2004a}. This indicates how a systematic 
improvement of the one-layer approximation could be achieved
by inclusion of more polymerization cycles, which will leave 
the result (\ref{Fs}) for the stall force 
unaffected.

Experimentally, force-velocity curves have been measured 
\cite{dogterom1997,Janson2004} but the stall force is not 
directly accessible. Nevertheless, an exact result such as 
 (\ref{Fs}) can help to  constrain the analysis of 
 experimental data.

We conclude this section with a short discussion of the effects of 
GTP-hydrolysis on the stall force of MTs. 
GTP-tubulin attaches and can hydrolyze within the MT 
to GDP-tubulin, which gives rise  to 
different off-rates $k_{\rm off,T}$ and 
$k_{\rm off,D}$  for GTP and GDP
monomers, respectively. This leads to a coupling between polymerization 
cycles and hydrolysis, if the probability $p_D$  
that a monomer at the filament end is of GDP-type becomes non-zero. 
For small probabilities $p_D$, effects from hydrolysis 
 can be included approximately 
by using an effective off-rate 
$k_{\rm off, eff} = p_T k_{\rm off,T}+p_Dk_{\rm off,D}$ in the result 
(\ref{Fs}) for the stall force, where $p_T=1-p_D$
is the probability that a monomer at the filament end is of 
GTP-type.
Because hydrolysis destabilizes the filament and 
$k_{\rm off,T}<k_{\rm off,D}$, hydrolysis will generally
{\em reduce}  the stall force. 
For single actin filaments, 
the effect of hydrolysis on the  force-velocity relation 
has been calculated in ref.\ \cite{Ranjith2009}. 
Hydrolysis and a non-zero $p_D$ also give rise to catastrophes 
as soon as the entire GTP-cap 
of a MT becomes hydrolyzed.

\begin{figure*}
\begin{center}
  \epsfig{file=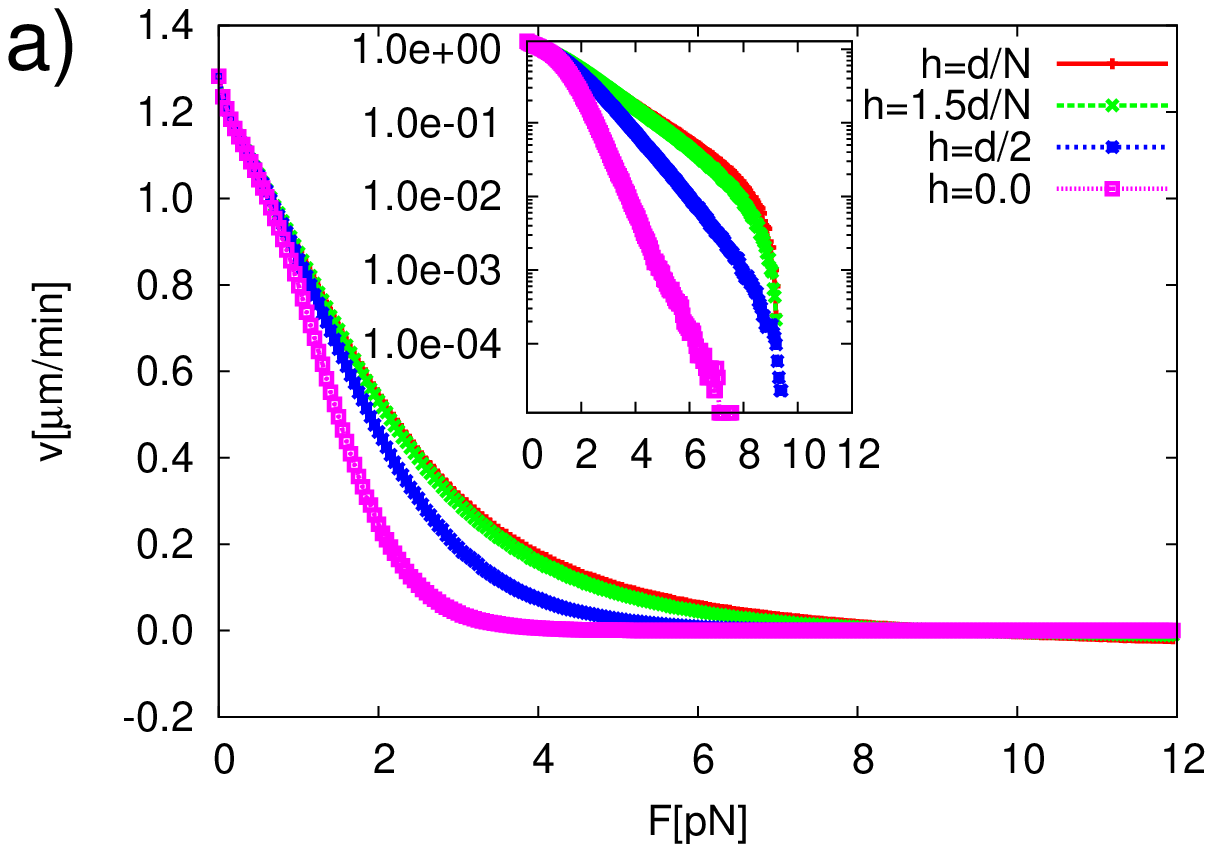,width=0.29\textwidth}~
  \epsfig{file=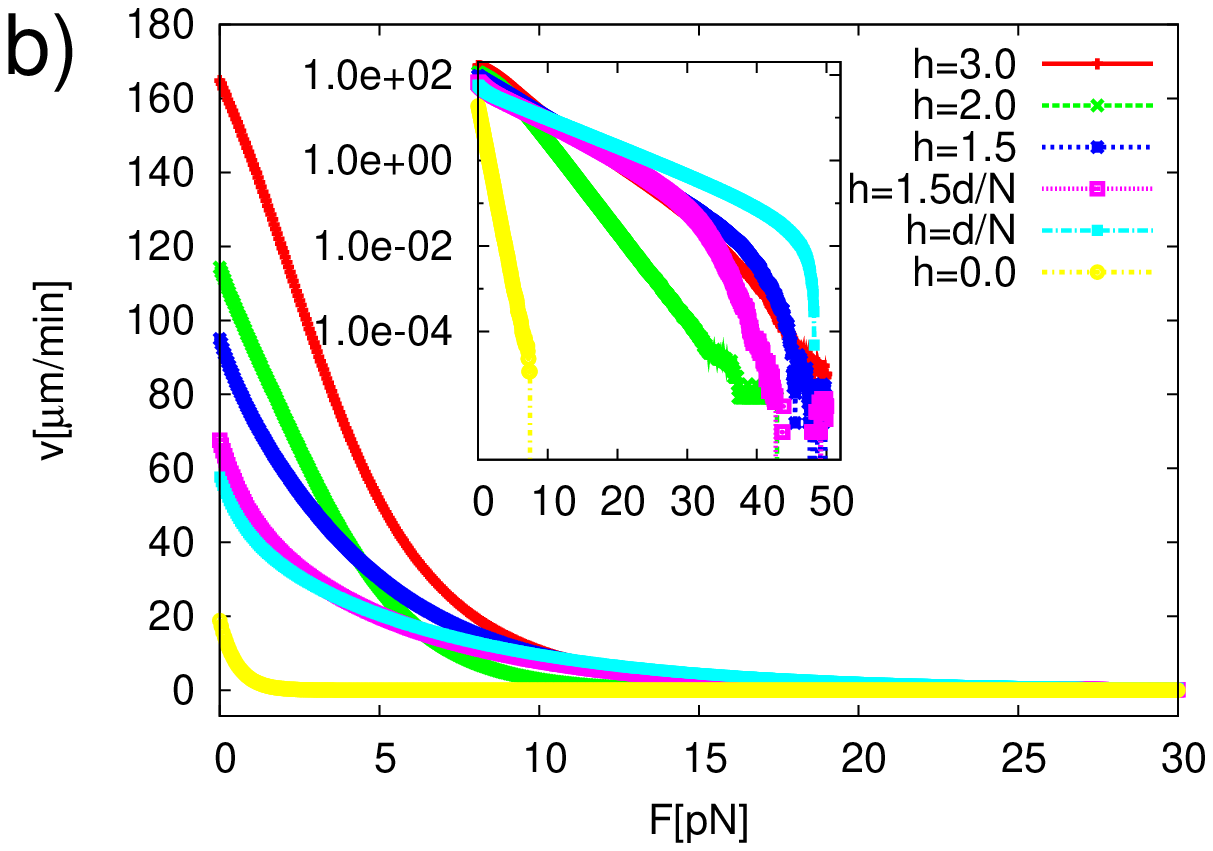,width=0.29\textwidth}~
  \epsfig{file=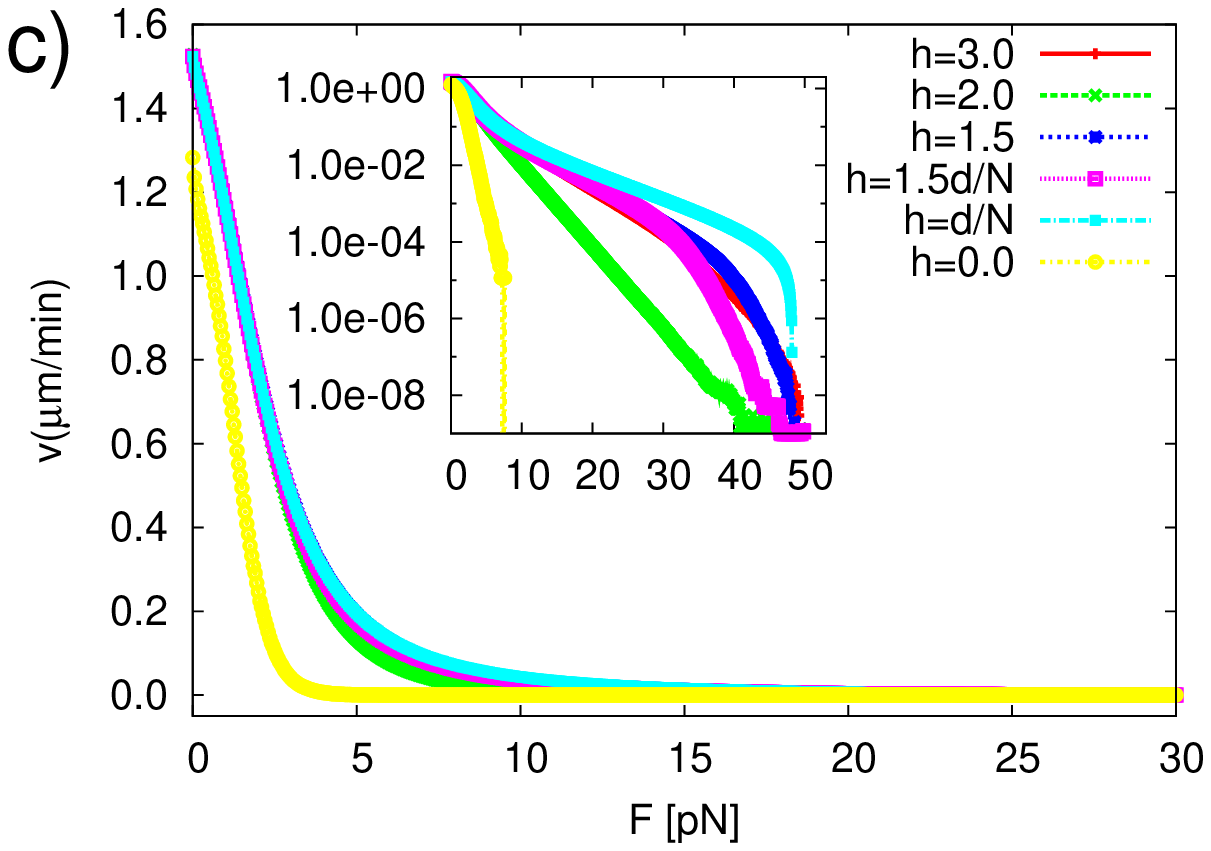,width=0.29\textwidth}
 \caption{Geometry-dependence of the force-velocity relation 
 for   $N=13$ PFs and two different values of the 
 lateral interaction, (a) $\eb=0 pN$ 
  (curves are independent of 
  the energy distribution factor $\theta_l$) 
   and (b),(c) $\eb=3.0 pN$, ((b) $\theta_l=0$, (c) $\theta_l=1$). 
  For each lateral interaction, 
   different geometry parameters $h$ (in units of nm and for
   monomer size  $d= 8 {\rm nm}$) are compared. 
  Insets are logarithmic plots.
 \label{pic_flat}
}
\end{center}
\end{figure*}

\section{Simulation results}

In simulations we can explore not only the stall force
but the full force-velocity relation. 
We characterize the growth process by the average growth 
velocity $v=\langle \dot x\rangle$.
We determine  the force-velocity relation as a function of 
{\em both} the  PF interaction (per length)
$\eb$ {\em and} the 
 geometry parameter $h$.  
From the force-velocity curves, we determine the stall 
force numerically 
and investigate how the stall force depends both on 
PF interactions and geometry parameter.
Some of our simulation results have also been obtained 
in ref.\ \cite{Tanase2004} using a fixed time step Monte-Carlo 
algorithm.

\subsection{Force-velocity relation}

We first describe results for the force-velocity 
relation of  $N=13$ PFs corresponding to a MT
using the rates (\ref{eq_onr1}).
The shape of the force-velocity relation depends on the 
PF interaction $\eb$ \cite{stukalin2004a,son2005}.
For $\eb=0$
all force-velocity curves  end at the same velocity   
 $v(0)= d(\ko -\kf)$ for zero force,
 independently of the geometry parameter $h$, 
  see fig.~\ref{pic_flat}a.
Simulations confirm that
 force-velocity curves for different geometry parameters $h$ 
 exhibit the same stall 
force $\Fs = N (k_BT/d) \ln(\ko/\kf)$ as predicted  in 
eq.\ (\ref{Fs}). 
The shape of the 
force-velocity curves between $F=0$ and the stall force $F=\Fs$, 
however, depends on the geometry 
parameter $h$, as shown in  fig.~\ref{pic_flat}a:
for forces $F>0$ and small $h$,  the kinetically limiting step is the 
insertion of the first monomer to an almost flat configuration. 
This rate-limiting step becomes faster for increasing $h$ because the 
increase $\Delta x$ of the leading tip becomes smaller.  
This results in steeper force-velocity curves for decreasing values 
of the geometry parameter $h$, as can be seen in fig.~\ref{pic_flat}a.

For  $\eb>0$, also the velocities at zero force $v(0)$ depend on the 
 geometry parameter $h$ and on the energy distribution factor 
$\theta_l$. For increasing $h$, 
the zero-force velocity $v(0)$ increases 
because the attractive PF interaction accelerates growth 
by reducing  the off-rate (for $\theta_l=1$, see fig.~\ref{pic_flat}(c))
 or increasing the 
on-rate (for $\theta_l=0$, see fig.~\ref{pic_flat}(b)).
Increasing the on-rate ($\theta_l=0$) leads to a much stronger effect. 
On the other hand, the simulations confirm  that the stall force 
remains identical  for different geometry parameters $h$ 
and for the different energy distribution factors 
$\theta_l$ in fig.~\ref{pic_flat}(b) and 
$\theta_l=0$ in  fig.~\ref{pic_flat}(c)
as predicted by eq.\ (\ref{Fs}).
Nevertheless, it is not possible to simply conclude that 
force-velocity curves become increasingly steep for increasing 
geometry parameter $h$. 
Only for $h=0$, strong suppression of 
the rate-limiting first insertion step by force always 
 results  in the steepest force-velocity 
curves, see  fig.\ \ref{pic_flat}. 
Then, the stall force is hard to 
determine because the force-velocity curve is extremely 
flat with measured zero velocity over a range of 
 higher forces.

\begin{figure}
\begin{center}
\epsfig{file=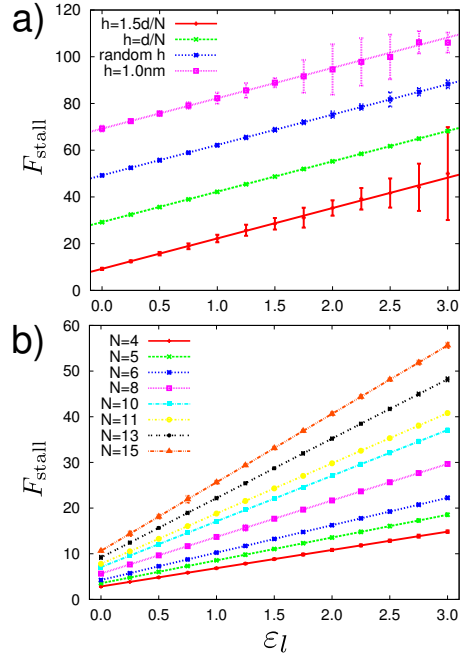,width=0.33\textwidth}
 \caption{ 
Stall force (in pN) as a function of lateral interaction energy  
$\eb$ (in pN) for $\theta_l=1$. Points represent simulation results, lines 
the analytical result (\ref{Fs}).
(a)  For $N=13$ PFs and different 
  geometry parameters $h=\frac{3}{2}d/N\approx 0.934 {\rm nm}$,
 $h=d/N\approx 0.615 {\rm nm}$, random $h$, $h=1.0 {\rm nm}$ 
(from bottom to top).
Curves 
 are shifted by $0$, $20$, $40$ and $60\,{\rm nm}$ respectively.  
(b) For a geometry parameter 
$h=d/N$ and different  PF numbers
$N=4$,$5$,$6$,$8$,$10$,$11$,$13$ and $15$ (from bottom to top).
 \label{ener_dep}}
\end{center}
\end{figure}

\subsection{Stall force}

It is possible to directly check the above  
exact analytical result (\ref{Fs})  for 
the stall force in  simulations 
by the condition $v(\Fs)=0$, i.e., by determining the force 
where the average growth velocity in the simulation vanishes. 
In simulations this is done by applying a linear interpolation 
to data points of force-velocity curves in vicinity of the stall 
force. 

 In fig.~\ref{ener_dep} we show the simulation results for the stall force 
as a function of the lateral interaction $\eb$ for (a) $N=13$ 
and  different values of the geometry parameter $h$ 
and for (b) fixed $h$ and  different values of  $N$.
The simulation results clearly show a linear increase with $\eb$ and $N$ 
and confirm  the analytical result  (\ref{Fs}). In particular, 
we find no dependence on the geometry parameter $h$. 

The derivation of (\ref{Fs}) was not limited to PF 
arrangement with a {\em constant} offset $h$ between neighboring 
PFs but predicts the same stall force for  completely arbitrary
arrangements of PFs. We checked this prediction 
by simulations of several random arrangements with random 
displacements between neighboring filaments, see  fig.~\ref{ener_dep}a
(green line).

\begin{figure}
\begin{center}
  \epsfig{file=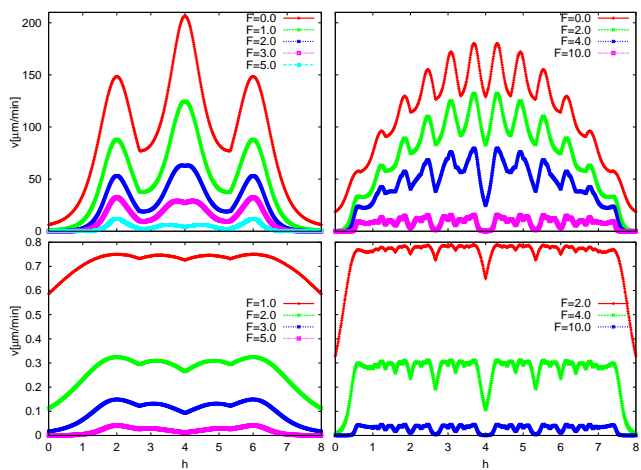,width=0.48\textwidth}
  \caption{Growth velocity as a function of  the 
    geometry parameter $h$ (in nm and for
   monomer size  $d= 8 {\rm nm}$) for
    $N=4$ (left) and $N=13$ (right) PFs for different 
    values of the force $F$ (in pN). The lateral interaction energy 
     is $\eb=3.0 {\rm pN}$ and 
     the  lateral energy distribution factor  $\theta_l=1$
     for both   top pictures 
      and $\theta_l=0$ for both bottom pictures. 
    For vanishing force $F=0$ the curves exhibit $N-1$ 
    maxima at $\bar{h} = i/N$ with $i=1,..,N-1$.  
     With increasing force, some maxima vanish and new maxima 
   can emerge. 
    }
  \label{diff_N4}
\end{center}
\end{figure}

\subsection{Dependence of velocity  on the geometry parameter}

The simulations allow us to explore how the  
growth velocity $v$ depends 
 on the geometry parameter $\bar{h}$ ($0<\bar{h}<1$) 
for different load forces $F$, 
see fig.~\ref{diff_N4}.
The curves $v(\bar{h})$ are symmetric with respect to the axis $\bar{h}=1/2$. 
For $\eb>0$ and  zero load force $F=0$, the 
curves $v(\bar{h})$  exhibit $N-1$ maxima corresponding to
relative displacements $\bar{h}=i/N$ with $i=1,..,N-1$.
At these values of $\bar{h}$, the relative displacement $h$ is commensurate 
with the monomer size $d$ such that polymerization cycles 
are possible, where all subsequently attached monomers gain the same 
lateral interaction energy. This avoids rate-limiting 
attachment steps and leads to optimal polymerization velocities. 
The maxima are very pronounced for energy distribution factors 
$\theta_l=0$, where the on-rate is exponentially increased 
by  lateral interactions and rather broad plateaus for $\theta_l=1$.
This implies that MT models using 
$\bar{h} = 1/13$ \cite{vanDoorn2000,Kolomeisky2001} overestimate 
the growth velocity as compared to the actual three-start helix 
with  $\bar{h} = 1.5/13$.

With increasing load $F$ the height of the maxima decreases, 
maxima can vanish or  become minima, 
and new local maxima can appear.     
Upon approaching the stall force $F\approx \Fs$, 
all curves $v(\bar{h})$
become flat, which supports 
the analytical result of an $h$-independent  stall force.

\section{Conclusion}

Based on the concept of polymerization cycles
we obtained the  exact result (\ref{Fs}) for the stall force
of polymerizing 
ensembles of rigid protofilaments with lateral 
interactions.
The stall force is a linear function of the number $N$ of filaments in the 
ensemble and increases linearly with the lateral interaction.
On the other hand, the stall force 
is independent of the geometry of the ensemble and load or energy 
distribution factors.
These results have been confirmed by simulations using the Gillespie 
algorithm. 
Simulations also show that the shape of the force-velocity relation
exhibits a pronounced dependence on the ensemble geometry
below the stall force.    
Our results are relevant for the interpretation of 
 experimental data on the force-velocity relation in
microtubule polymerization and 
in the polymerization of bundles of interacting 
actin filaments or microtubules.




\begin{thebibliography}{99}

\bibitem{bray}
  \Name{Bray D.}
  \Book{Cell Movements: From Molecules to Motility}
  \Publ{Garland Publishing}
  \Year{2001}

\bibitem{mogilner_rev}
  \Name{Mogilner A.}
  \REVIEW{Curr. Opin. Cell Biol.}{18}{2006}{32}.

\bibitem{dogterom1997}
   \Name{Dogterom M. \and Yurke B.}
     \REVIEW{Science}{278}{1997}{856}.

\bibitem{Hill1987}
    \Name{Hill T.L.}
  \Book{Linear aggregation theory in cell biology}
  \Publ{Springer}
  \Year{1987}

\bibitem{faix06}
  \Name{J.Faix \and K.Rottner}
  \REVIEW{Curr. Opin. Cell Biol.}{18}{2006}{18}.

\bibitem{footer07}
     \Name{Footer M.J.,Kerssemakers J.W.J., Theriot J.A. \and 
     Dogterom M.}
     \REVIEW{Proc. Nat. Acad. Sci. USA}{104}{2007}{2181}.

\bibitem{Janson2004}
   \Name{Janson M. \and Dogterom M.}
     \REVIEW{Phys. Rev. Lett.}{92}{2004}{248101}.


\bibitem{peskin1993a}
	\Name{Peskin C.S., Odell G.M. \and Oster G.F.}
	\REVIEW{Biophys. J.}{65}{1993}{316}.


\bibitem{ostermogilner1999} 
		\Name{Mogiler A. \and Oster G.}
		\REVIEW{Eur. Biophys. J.}{28}{1999}{235}. 



\bibitem{vanDoorn2000} 
		\Name{van Doorn G.S., T\v{a}nase C., Mulder B.M. \and 
                  Dogterom M.}
		\REVIEW{Eur. Biophys. J.}{29}{2000}{2}.

\bibitem{Kolomeisky2001}
        \Name{Kolomeisky A.B. \and Fisher M.E.}
        \REVIEW{Biophys. J.}{80}{2001}{149}.


\bibitem{mahadevan00}
  \Name{L.Mahadevan \and P.Matsudaira}
  \REVIEW{Science}{288}{2000}{95}.

\bibitem{kuehne09}
          \Name{K{\"u}hne T., Lipowsky R. \and Kierfeld J.}
           \REVIEW{EPL}{86}{2009}{68002}.

\bibitem{Tanase2004}
  \Name{Tanase C.}
  \Book{Physical Modeling of Microtubule 
  Force Generation and Self-Organization,}
  \Year{PhD Thesis, Wageningen University, 2004}.

\bibitem{stukalin2004a} 
		\Name{Stukalin E.B. \and Kolomeisky A.B.}
		\REVIEW{J. Chem. Phys.}{121}{2004}{1097}.


\bibitem{son2005} 
		\Name{Son J., Orkoulas G. \and Kolomeisky A.B.}
		\REVIEW{J. Chem. Phys.}{123}{2005}{124902}.

\bibitem{nogales2001}
             \Name{Li H., DeRosier D.J., Nicholson W.V.,  Nogales E. \and 
                 Downing K.H.} 
           \REVIEW{Structure}{10}{2002}{1317}.

\bibitem{KKL05}
  \Name{Kierfeld J., K{\"u}hne T. \and Lipowsky R.}
 \REVIEW{Phys. Rev. Lett.}{95}{2005}{038102}.

\bibitem{Gov2008}
 \Name{Gov N.}
 \REVIEW{Phys. Rev. E}{78}{2008}{011916}.

\bibitem{Yang2010}
 \Name{Yang Y., Meyer R.B. \and Hagan M.F.}
 \REVIEW{Phys. Rev. Lett.}{104}{2010}{258102}.

\bibitem{Grason2010}
  \Name{Grason G.}
 \REVIEW{Phys. Rev. Lett.}{105}{2010}{045502}.

\bibitem{chretien1991} 
		\Name{Chretien D. \and Wade R.H.}
		\REVIEW{Biol. Cell}{71}{1991}{161}. 

\bibitem{gillespie1977} 
		\Name{Gillespie D.T.}
		\REVIEW{J. Chem. Phys.}{28}{1977}{395}. 

\bibitem{hill1966}
    \Name{Hill T.L.}
    \REVIEW{Journal of theoretical biology}{10}{1966}{442}.

\bibitem{schnakenberg1976}
     \Name{Schnakenberg J.}
    \REVIEW{Rev. Mod. Phys.}{48}{1976}{571}.

\bibitem{hill1989}
       \Name{Hill T.L.}
  \Book{Free energy transduction and biochemical cycle kinetics}
  \Publ{Springer}
  \Year{1989}


\bibitem{Wegscheider1901}
     \Name{Wegscheider R.}
  \REVIEW{Z. Phys. Chem.}{39}{1901}{257}. 

\bibitem{Ranjith2009}
  \Name{Ranjith P., Lacoste D., Mallick K. \and Joanny, J.-F.}
 \REVIEW{Biophys. J.}{96}{2009}{2146}.

\end{thebibliography}
\end{document}